\documentclass{acm_proc_article-sp}

\usepackage{amssymb}
\usepackage{url}
\usepackage{amsmath}
\usepackage{graphicx}
\usepackage{fancyvrb}

\usepackage[linesnumbered,ruled,vlined]{algorithm2e}

\usepackage{chngpage}

\usepackage{relsize}
\usepackage{makeidx} 
\usepackage{multirow}

\def\eg{{\em e.g.}}
\def\ie{{\em i.e.}}
\newcommand{\hide}[1]{}

\begin{document}
%

\title{ProbReach: Verified Probabilistic Delta-Reachability for Stochastic Hybrid Systems}
%
%
%
%
%

\numberofauthors{2} 
%
\author{
%
%
\alignauthor
Fedor Shmarov\\
       \affaddr{School of Computing Science}\\
       \affaddr{Newcastle University}\\
       \affaddr{Newcastle upon Tyne, UK}\\
       \email{f.shmarov@ncl.ac.uk}\\
\alignauthor
Paolo Zuliani\\
       \affaddr{School of Computing Science}\\
       \affaddr{Newcastle University}\\
       \affaddr{Newcastle upon Tyne, UK}\\
       \email{paolo.zuliani@ncl.ac.uk}\\
}

\maketitle
\begin{abstract}
We present ProbReach, a tool for verifying probabilistic reachability 
for stochastic hybrid systems, i.e., computing the probability that the system
reaches an unsafe region of the state space. In particular, ProbReach 
will compute an arbitrarily small interval which is guaranteed to contain 
the required probability. Standard (non-probabilistic) reachability
is undecidable even for linear hybrid systems. In ProbReach we adopt 
the weaker notion of delta-reachability, in which the unsafe region is
overapproximated by a user-defined parameter (delta). This choice leads 
to false alarms, but also makes the reachability problem decidable for
virtually any hybrid system. In ProbReach we have implemented a probabilistic
version of delta-reachability that is suited for hybrid systems whose 
stochastic behaviour is given in terms of random initial conditions.
In this paper we introduce the capabilities of ProbReach, give an overview 
of the parallel implementation, and present results for several benchmarks 
involving highly non-linear hybrid systems.
\end{abstract}




\section{Introduction}
In modern society, we interact with cyber-physical systems (\eg, cars and airplanes) on a daily
basis. Some of these systems are safety-critical, with human lives crucially depending on their 
reliability and correctness. Thus, verification of cyber-physical systems is extremely important.

Verifying cyber-physical systems is a very difficult task and can be performed in various ways. We employ hybrid systems as an expressive framework for modelling and verification of cyber-physical systems. One of the most important properties investigated by researchers in hybrid systems is reachability.
The main reason being that many verification problems can be presented as reachability problems. In other words, we wish to verify whether a hybrid system reaches an {\em unsafe} region --- a subset of the state space of the system representing an unwanted behaviour. The reachability problem is undecidable 
in general (even for linear hybrid systems \cite{DBLP:conf/hybrid/AlurCHH92}). We avoid 
undecidability issues by solving instead the weaker $\delta$-reachability problem 
\cite{DBLP:journals/corr/GaoKCC14}, which asks whether a hybrid system reaches an 
{\em overapproximation} of the unsafe region.

In this paper we focus on hybrid systems featuring stochastic behaviour. Such systems frequently
arise when modelling real-world cyber-physical systems. For example, random behaviour can happen 
due to soft errors in some components of the system. Without a doubt this can cause the whole system behaving in a faulty way. By investigating a problematic component, its characteristics (\eg, error distribution) can be obtained. In this case it might be necessary not only to predict an undesired behaviour but also show that the probability of occurrence of a bad event is below (or above) some required threshold. This problem is called probabilistic reachability, and it can be expressed for stochastic hybrid systems. In particular, we consider hybrid systems with random continuous/discrete initial parameters. Such parameters are assigned in the initial mode and remain unchanged throughout the system's evolution. Having a probability measure on random parameters we can assess quantitative properties of hybrid systems such as the probability of reaching an {\em unsafe} set of states.

We implemented the tool {\em ProbReach} which performs verified computation of the probability 
that a hybrid system reaches an unsafe region within a finite number of discrete steps. In particular,
our tool implements a general procedure for computing an interval of arbitrarily small length 
which is {\em guaranteed} to contain the {\em exact} value of the probability. {\em ProbReach} 
works for general hybrid systems whose continuous dynamics is given, \eg, as a solution of 
ordinary differential equations (ODEs). 
Our tool uses $\delta$-complete decision procedures \cite{DBLP:conf/cade/GaoKC13} and implements 
a verified integration procedure \cite{Petras:VerifiedNI} used for integrating probability measures of random variables.

\paragraph{\bf \em Related work}
To the best of our knowledge, SiSAT \cite{Sisat} is the only tool that can perform verified reachability analysis in hybrid systems with random parameters. However, it supports only discrete random variables, while {\em ProbReach} accepts continuous and discrete random initial parameters.
A recent work \cite{Franzle:STTT14} proposes a statistical model checking technique for verifying
hybrid systems with continuous nondeterminism, thereby significantly expanding the class of systems
analysable. However, the approach is based on statistical planning algorithms from AI, and therefore
it cannot offer the absolute guarantees provided by {\em ProbReach}. A similar approach has been
taken by the SReach tool \cite{SReach}, which combines statistical techniques with $\delta$-complete 
procedures. The advantages of SReach are its ability to handle large numbers of initial random variables
and probabilistic transitions. Again, SReach can only offer statistical guarantees, while
{\em ProbReach} focuses on absolutely correct results. Also, in Section \ref{sec:Exp} 
we essentially show that {\em ProbReach} can be as fast as statistical (Monte Carlo) methods.

In this paper we explain the theoretical background of {\em ProbReach}, its implementation details 
and consider several case studies such as an insulin glucose regulatory system 
\cite{Sankaranarayanan:2012:SII:2415548.2415569}, a controlled bouncing ball \cite{ADHS09}, and 
a thermostat model.

\section{Background}
We give here a brief overview of the theory underlying {\em ProbReach}. For simplicity
we focus on one continuous random parameter only --- more details can
be found in \cite{ProbDeltaReach}. {\em ProbReach} addresses the following problem: 
\begin{quote}
{\em what is the probability that a hybrid system with random initial parameters reaches the 
unsafe region $U$ in $k$ steps?}
\end{quote}
As this problem is in general undecidable, we adopt the weaker notion of $\delta$-reachability.
In our setting it means that {\em ProbReach} will actually compute an interval of a user-specified 
length $\epsilon >0$ that is guaranteed to contain the reachability probability.
The main idea of the approach implemented in the tool is to compute the probability by integrating an indicator function over the probability measure of the random variable as:
\begin{equation*}
\int_{\Omega} I_{U}(r)dP(r)
\end{equation*}
where $P(r)$ is a probability measure of the random variable, $\Omega$ is the domain of the 
random variable, and $I_{U}$ is the indicator function defined as:
\begin{equation*}
        I_{U}(r) = 
        \begin{cases}
                1, \text{system with parameter $r$ reaches $U$ in $k$-steps} \\
                0, \text{otherwise.}
        \end{cases}
\end{equation*}
The procedure for solving probabilistic reachability combines a validated integration procedure and a decision procedure. The first one integrates a probability measure (probability density function) of a random variable and obtains a partition of the random variable domain which guarantees that the probability interval is not larger than the desired length $\epsilon$. The second procedure evaluates the indicator function on each of the intervals from the obtained partition and performs a partial analysis of the interval if necessary.

\paragraph{\bf \em Validated Integration Procedure}
The problem here is to compute the integral function defined by
\[
\mathcal{I}([a,b])=\int_a^b f(x) dx 
\]
up to an error $\epsilon$. 
In the implementation of our validated integration procedure we employ the (1/3) Simpson rule which,
by applying interval arithmetics \cite{ValidatedIntegration}, can be formulated as:
\[
\begin{split}
\mathcal{I}([a,b]) \in [\mathcal{I}]([a, b]) = \frac{b-a}{6}([f](a) + 4 [f](\frac{a+b}{2})) + \\
[f](b)) - \frac{(b-a)^5}{2880}[f]^{(4)}([a,b])
\end{split}
\]
where $[\mathcal{I}]$ and $[f]$ are the interval extensions of functions $\mathcal{I}$ and $f$. 
Then by the definition of integral:
\[
\begin{split}
\mathcal{I}([a,b]) \in \Sigma_{i = 1}^{n} [\mathcal{I}]([r]_{i})
\end{split}
\]
where $n$ is a number of disjoint intervals $[r]_{i}$ that partition $[a, b]$. Interval extensions 
can be readily computed using interval arithmetics libraries such as FILIB++ \cite{filib}.

\paragraph{\bf \em Decision Procedure}
Our decision procedure encodes bounded $\delta$-reachability in hybrid systems as a first-order logic formula. This formula is then passed to a $\delta$-complete decision procedure 
\cite{DBLP:conf/cade/GaoAC12} which uses the notion of $\delta$-weakening of a logical formula. 
Basically, the main idea is to perform evaluation of a weaker (decidable) formula and make a conclusion about the initial formula on this basis. Given an arbitrary first order formula the $\delta$-complete procedure returns \textbf{unsat} if the formula is false and \textbf{$\delta$-sat} if its weakening is true. Hence, unlike \textbf{unsat}, \textbf{$\delta$-sat} is a {\em weak} answer as it does not imply the satisfiability of the formula. We use this fact to define a decision procedure for verifying the indicator function above. 
The decision procedure comprises two formulas $\phi$ and $\phi^{C}$ which are defined as following: 
\begin{itemize}
	\item{$\phi([r]_{i})$ is {\bf true} if the interval $[r]_{i}$ contains a value $r$ such 
		that $I_U(r) = 1$ and {\bf false} if $I_U(r) = 0$ for all the points of the interval}
	\item{$\phi^{C}([r]_{i})$ is {\bf true} if there is a value in $[r]_{i}$ such that $I_U(r) = 0$ and {\bf false} if $I_U(r) = 1$ everywhere on the interval}.
\end{itemize}

Verifying now both formulas using dReach\footnote{\url{http://dreal.cs.cmu.edu/dreach.html}}, we obtain four 
outcomes which can be interpreted as follows:
\begin{itemize}
	\item{$\phi([r]_{i})$ is \textbf{unsat}. Hence, $I_{U}(r) = 0$ in all points on $[r]_{i}$ {\em for sure}.}
	\item{$\phi([r]_{i})$ is \textbf{$\delta$-sat}. Then there is a value in the interval $[r]_{i}$ such that the system reaches the unsafe region $U$ or its weaker definition ($\delta$-weakening).}
	\item{$\phi^C([r]_{i})$ is \textbf{unsat}. Thus, $I_{U}(r) = 1$ point-wise on $[r]_{i}$ {\em for sure}.}
	\item{$\phi^C([r]_{i})$ is \textbf{$\delta$-sat}. Then there is a value in the interval $[r]_{i}$ such that the system stays outside the unsafe region or its weakening within the $k$-th step.}
\end{itemize}
As it was stated above, only \textbf{unsat} returned for either of the formulas guarantees the correctness of the interval validation. Therefore, if both formulas evaluates as \textbf{$\delta$-sat} then either a {\em false alarm} is obtained (when a formula which should be unsatisfiable is verified as $\delta$-sat because of a relatively large value of $\delta$ used for verification) or the analysed 
interval is {\em mixed} (\ie, it contains a value $r$ for which $I_{U}(r) = 0$ and also a value $s$
for which $I_{U}(s) = 1$) which means that the interval should be partitioned and verified again.
The pseudo-code of the algorithm implemented in {\em ProbReach} is presented in 
Algorithm \ref{alg:prob-reach}.

\section{System overview}

\begin{figure}[ht!] 
\centering
\includegraphics[width=70mm]{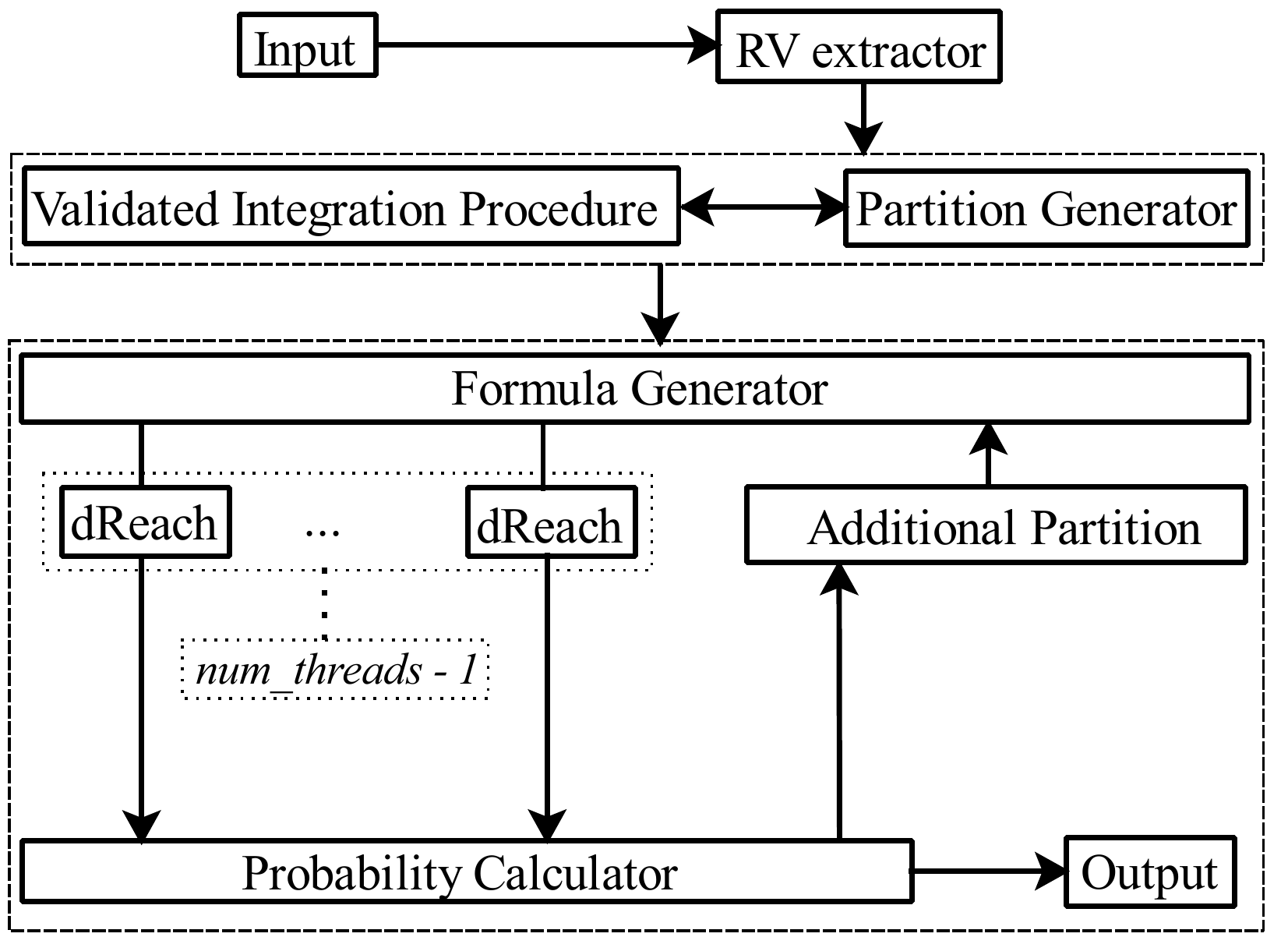}
\caption{Architecture of ProbReach}
\label{fig:architecture}
\end{figure}

This section aims giving an overview of the main components of {\em ProbReach}, 
their interaction, and implementation details. The architecture of the tool is shown in Figure \ref{fig:architecture}.

\begin{algorithm} \label{alg:prob-reach}

\DontPrintSemicolon
\SetKwComment{tcz}{\{}{\}}
\SetKwInOut{Input}{Input}\SetKwInOut{Output}{Output}

\Input{probability density $f$, $t\in(0,1)\cap\mathbb{Q}$, $\epsilon\in (0,1]\cap\mathbb{Q}$, formula $\phi,\phi^C$}
\Output{interval $[I]$: $\int_B f \in [I]$ and $width([I]) \leq \epsilon$}

$\epsilon_{inf} = t \epsilon$\;
$\epsilon_{prob} = (1-t) \epsilon$\;
$[a, b] = bounds(f, \epsilon_{inf})$\tcz*{obtain bounds}
$B.push(integral(f, [a, b], \epsilon_{prob}))$\tcz*{get partition}
$[P_{lower}] = [0.0, 0.0]$\tcz*{interval for lower approx}
$[P_{upper}] = [1.0, 1.0]$\tcz*{interval for upper approx}
\While{$\overline{[P_{upper}]} - \underline{[P_{lower}]} > \epsilon_{prob}$} { \label{alg:main-loop}
	$D = \emptyset$\tcz*{extra interval divisions}
	\While{$size(B) > 0$} { \label{alg:partition-verification}
	$\{[x], [S]([x])\} = B$.pop()\tcz*{get an interval}
	\uIf(\tcz*[f]{call dReach}){$\phi([x])$ == $\delta$-sat} {
		\uIf(\tcz*[f]{call dReach}){$\phi^{C}([x])$ == $\delta$-sat} {
			$D$.push($\{[\underline{x}, mid([x])], 
				[S([\underline{x}, mid([x])])]\}$)
			$D$.push($\{[mid([x]), \overline{x}], 
				[S([mid([x]), \overline{x})]\}$)
		}\lElse {$[P_{lower}] = [P_{lower}] + [S]([x])$\tcz*[f]{update int}}
	} \lElse {$[P_{upper}] = [P_{upper}] - [S]([x])$\tcz*[f]{update int}}
	}
	$B = D$\; \label{alg:extra-partition}
}
$[P_{upper}] = [P_{upper}] + 1 - \int_{a}^{b}\, f(x)\, dx $\tcz*[f]{add leftovers}

\Return $[\underline{[P_{lower}]}, \overline{[P_{upper}]}]$\;

\caption{{\em ProbReach} (one cont.~random parameter)}
\end{algorithm}

\paragraph{\bf \em Input}
In the first step {\em ProbReach} validates the input and extracts all the necessary data. The application requires a single input file (containing $\phi$ and $\phi^{C}$) in \texttt{PDRH} format. This file is used further as templates by the {\bf Formula Generator}. An example of the \texttt{PDRH} model of a two-mode thermostat is given below. Note in particular
the declaration of a random parameter \texttt{x} distributed as a normal with mean 30 and standard 
deviation 1. 
\begin{Verbatim}[frame=single, numbers=left, samepage = true]
#define K 1.0
[0, 5] time;
[0, 1000] tau;
//random parameter declaration
N(30, 1) x;	
//cooling mode
{ mode 1;       
invt:
(x >= 18);
flow:
d/dt[x] = - x * K;
d/dt[tau] = 10.0;
jump:
(x <= 18) ==> @2 (and (x' = x) (tau' = tau));
}
//heating mode
{ mode 2;       
invt:
(x <= 22);
flow:
d/dt[x] = - K * (x - 30);
d/dt[tau] = 10.0;
jump:
(x >= 22) ==> @1 (and (x' = x) (tau' = tau));
}
//initial state
init:           
@1(and (tau = 0));
//unsafe region
goal:           
@2(and (x >= 19.9) (x <= 20.1) (tau = 6));
//unsafe region complement
goal_c:           
@2(or (x < 19.9) (x > 20.1) (tau = 6));
\end{Verbatim}
The details of how to use {\em ProbReach} are given in {\bf \em Application Usage} section.

The aim of the {\bf RV extractor} is to read all the random variables from the model file containing $\phi$, ignoring any other parameter declarations. 
The tool recognises most of the frequently used distributions (\eg, uniform, normal, exponential), and once the random variables are successfully extracted, their probability density function is automatically generated. Hence, {\em ProbReach} is not restricted to some set of predefined random variables and can be extended to allow user-defined distributions (by simply providing a probability density function).

\paragraph{\bf \em Verified integration and Partition generation}
Many useful random variables are defined over unbounded intervals (\eg, normal distribution). However, it was shown in the previous section how to perform verified integration and reachability analysis 
over bounded intervals only. We cope with unbounded intervals by making a trade-off. Given a desired length $\epsilon$ of the probability interval we choose a value $t \in (0, 1)$ (can be also defined by the user) and obtain an interval $[a, b]$ such that:
\begin{equation*}
\int_{a}^{b}\, f(r)\, dr > (1-t)\epsilon
\end{equation*}
Finding $a$ and $b$ can be actually encoded as a logical formula which can be solved by 
dReal \cite{DBLP:conf/cade/GaoKC13}.

The intuition behind this is that we assume that the indicator function equals to 1 outside the interval $[a, b]$. In case if it is not true (the indicator function is 0 in some points outside the considered bounded domain) the integral of the indicator function over the unbounded intervals will be still bounded by $t\epsilon$ (as the integral of a probability density function on interval $(-\infty, \infty)$ is 1). 

Then, the {\bf Validated Integration Procedure} computes a definite integral of the probability
density function on the obtained finite interval. This is achieved through an iterative partitioning (by {\bf Partition Generator}) of the integration domain until on each interval $[r]_{i}$ the value of the integral is enclosed by an interval of the length $(1 - t)\epsilon\frac{width([r]_{i})}{width([a, b])}$. For such a partition it is guaranteed that the value of the integral over the bounded domain belongs to an interval of length $(1 - t)\epsilon$.

\paragraph{\bf \em Partition verification}
Once the {\em correct} partition is obtained, each interval $[r]_{i}$ is used to generate two model files (encoding $\phi([r]_{i}$) and $\phi^{C}([r]_{i})$) in \texttt{DRH} format which are then verified by dReach. This routine was parallelised using the OpenMP shared memory library (see the code below).
\begin{Verbatim} [frame=single, numbers=left, samepage = true]
//setting a number of threads
int num_threads = omp_get_max_threads();
if (num_threads > 1)
{
    omp_set_num_threads(num_threads - 1);
}
//Algorithm 1 line 6 loop
{
   #pragma_omp_for
   {
        //Algorithm 1 line 8
   }
   //Algorithm 1 line 15
   while (B.size() < num_threads - 1)
   {
       //partition B to reduce CPU idle
   }
}
\end{Verbatim}
Initially, the application gets the maximum number of available cores (\texttt{num\_threads}) and uses \texttt{num\_threads - 1} (if more then one is available) of them to perform the computation leaving one core to let the computer executing background tasks. Then the partition is distributed between \texttt{num\_threads - 1} threads and each of them evaluates its interval with dReach.

Now, if for the analysed interval either of the formulas is \textbf{unsat} then {\bf Probability Calculator} modifies the probability bounds:
\begin{itemize}
	\item{if $\phi([r]_{i})$ is \textbf{unsat} then $[r]_{i}$ is used for calculating $P_{upper}$ (probability upper bound). The integral of the probability density over the interval $[r]_i$ is subtracted from $P_{upper}$; initially 
we of course have $P_{upper} = 1$.}
	\item{if $\phi^{C}([r]_{i})$ is \textbf{unsat} then $[r]_{i}$ is used for calculating $P_{lower}$ (probability lower bound). The integral of the probability density over the interval $[r]_i$ is added to $P_{under}$, starting initially with $P_{under} = 0$.}
\end{itemize}

However, both formulas may be evaluated as \textbf{$\delta$-sat} for a given interval from the partition. This suggests that either a {\em false alarm} is obtained or the interval is {\em mixed} (it contains values satisfying both formulas). Then, such an interval is subject to {\bf Additional Partition}, which should further undergo the described cycle once again. In the parallel implementation, all {\em mixed} intervals are partitioned until their number reaches \texttt{num\_threads - 1}, to reduce CPU idle time. Extra partitioning can be performed arbitrarily many times as it does not alter the correctness of the result. The described routine stops when the length of the interval $[P_{lower}, P_{upper}]$ is shorter than $(1 - t)\epsilon$.
Hence, taking into account the assumption about the value of the indicator function outside the bounded domain the probability is guaranteed to be contained inside the interval of the length $t\epsilon + (1 - t)\epsilon = \epsilon$.

Finally, we note that at {\em any} point in time during the computation, the {\em exact} value 
of the probability 
belongs to the interval $[P_{lower}, P_{upper}]$, which is written in output when the interval 
bounds change. This might be advantageous for time-critical verification scenarios, as 
the user can specify a computation timeout. Thus, despite the fact that the desired precision 
might not be achievable within the specified timeframe, the obtained result is still complete 
in the sense that the desired probability is {\em guaranteed} to be inside the computed interval.

\paragraph{\bf \em Implementation details}
{\em ProbReach} has been implemented in C++, using the CAPD library\footnote{\url{http://capd.ii.uj.edu.pl}} for interval operations. Input analysis is performed using the C++11 regular expression engine. Parallelisation of the code was achieved using OpenMP, and both versions of the tool (parallel and sequential) were built and tested. The parallel implementation running on 24 cores demonstrated a 8-10 times speed up in comparison to the sequential one. 
\paragraph{\bf \em Application usage}
Once the tool has been compiled, the executable is put into \texttt{\allowbreak<ProbReach-directory>/bin}. Then the tool can be called from the command line as \texttt{./ProbReach <options> <model-file.pdrh> -\--dreach <dReach-options> -\--dreal <dReal-options>}. The {\em ProbReach} options are specified below:
\begin{Verbatim} [samepage = true]
options:
-e <double> - length of probability interval 
	or max length of box edge (default 0.001)
-l <string> - path to dReach binary (default dReach)
-t <int> - number of CPU cores (default 1)
-h/--help - help message
--version - version of the tool
--verbose - output computation details
--dreach - delimits dReach options 
	(e.g., reachability depth)
--dreal - delimits dReal options 
	(e.g., precision, ode step)
\end{Verbatim}
\paragraph{\bf \em Tool availability}
The source code of {\em ProbReach} and installation instructions are available on \texttt{https://github.com\allowbreak /dreal/probreach}.
We also implemented a web application to display {\em ProbReach}'s results.
{\em ProbReach} outputs intermediate probability intervals to a \texttt{JSON} file which can be visualised by \texttt{\\\allowbreak https://homepages.ncl.ac.uk/f.shmarov/probreach/}.

\section{Experiments}\label{sec:Exp}
The description of all the models and verification scenarios are given in the Appendix. All the experiments were carried out on a Intel Xeon E5-2690 2.90GHz multi-core system running Linux Ubuntu 14.04LTS.
The parallel version of {\em ProbReach} ran on 24 cores. The results were also validated using a Monte Carlo method in MATLAB. We calculated confidence intervals using the sample size returned by the Chernoff-Hoeffding \cite{hoeffding} bound
$ N = \frac{\log{\frac{1}{1 - c}}}{2\zeta^2}$,
where $\zeta$ is the interval half-width and $c$ is the coverage probability. The results are
presented in Table \ref{table:results}. 
\vspace{-3ex}
\paragraph{\bf \em Results analysis}
In most of the experiments {\em ProbReach} demonstrated a better performance than the Monte Carlo method. However, for the Insulin-Glucose (IG) model the Monte Carlo method was faster for the two scenarios
considered. Nevertheless, reducing the length of the confidence interval causes a quadratic growth 
in the sample size. For example, obtaining a confidence interval of size $10^{-4}$ with 
coverage $0.999$ requires $1.3815510558\times10^{9}$ samples, with an estimated CPU time 
of $2.3 \times 10^{9}$ seconds. {\em ProbReach} computes a guaranteed enclosure of size smaller
than $10^{-4}$ in about $3.5 \times 10^{6}$ seconds. Hence, for stronger precisions 
(\ie, smaller $\epsilon$) {\em ProbReach} performs better than Monte Carlo method. 

Considering the results for the thermostat model (see rows T4(1.7) in Table \ref{table:results}), 
the Monte Carlo method returned a probability estimate (number of successes divided by number of samples) of $9.438088\times 10^{-8}$ with a relatively large confidence interval ($10^{-5}$) using
33,015 seconds of CPU time. {\em ProbReach} can compute an interval of size about $10^{-9}$ in just
268 seconds. Computing a confidence interval of length $10^{-9}$ with coverage $0.99999$ requires 
$2.3025850929\times10^{19}$ samples, which suggests that {\em ProbReach} can be very efficient 
for rare event verification.

\begin{table*}[t!] 
\begin{adjustwidth}{-1in}{-1in}
\centering
\begin{tabular}{|l |c |c |c |c |c |c |c |r |}
\hline\hline
{\bf Tool} & {\bf Model} & {\bf $k$} & {\bf $\epsilon$} & {\bf $length$} && {\bf Probability interval} & {\bf $CPU_{seq}$} & {\bf $CPU_{par}$}\\
\hline
\multirow{4}{3em}{Prob Reach}& \multirow{4}{1em}{BB} & 0 & $10^{-9}$ & 5.0e-10 && [8.21757e-05, 8.21762e-05] & 64 & 7\\ 
&& 1 & $10^{-9}$ & 1.0e-09 && [0.1379483631, 0.1379483641] & 192 & 29\\
&& 2 & $10^{-9}$ & 9.9e-10 && [0.50868960502, 0.50868960601] & 927 & 164\\
&& 3 & $10^{-9}$ & 8.0e-10 && [0.7387674005, 0.7387674013] & 3806 & 563\\
\hline
& & & {\bf $\zeta$} & {\bf $c$} & $P$ & {\bf Confidence interval} & {\bf $CPU_{seq}$} & {\bf Sample size}\\
\hline
\multirow{4}{3em}{Monte Carlo}& \multirow{4}{1em}{BB} & 0 & $5\cdot10^{-6}$ & 0.99999 & 8.220032e-05 & [7.720032e-05, 8.720032e-05] & 16,455 & 230,258,509,300\\
&& 1 & $5\cdot10^{-6}$ & 0.99999 & 0.1379449 &[0.1379399, 0.1379499] & 19,646 & 230,258,509,300\\ 
&& 2 & $5\cdot10^{-6}$ & 0.99999 & 0.5086939 &[0.5086889, 0.5086989] & 21,197 & 230,258,509,300\\ 
&& 3 & $5\cdot10^{-6}$ & 0.99999 & 0.7387684 &[0.7387634, 0.7387734] & 20,975 & 230,258,509,300\\ 
\hline
\hline
& & & {\bf $\epsilon$} & {\bf $length$} && {\bf Probability interval} & {\bf $CPU_{seq}$} & {\bf $CPU_{par}$}\\ 
\hline
\multirow{3}{3em}{Prob Reach}& T2(0.6) &1 & $10^{-9}$ & 9.46e-10 && [0.006678444555, 0.0066784456] & 71 & 7\\
&T2(1.8)&5 & $10^{-9}$ & 1.0e-9 && [0.0026170599, 0.0026170609] & 213 & 23\\ 
&T2(2.4)&7 & $10^{-9}$ & 1.0e-9 && [0.0015794358, 0.0015794368] & 364 & 49\\
\hline
& & & {\bf $\zeta$} & {\bf $c$} &$P$& {\bf Confidence interval} & {\bf $CPU_{seq}$} & {\bf Sample size}\\
\hline
\multirow{3}{3em}{Monte Carlo}& T2(0.6) &1 & $5\cdot10^{-6}$ & 0.99999 &0.006679496& [0.006674496, 0.006684496] & 31,822 & 230,258,509,300\\
&T2(1.8)&5 & $5\cdot10^{-6}$ & 0.99999 &0.002616634& [0.002611634, 0.002621634] & 33,287 & 230,258,509,300\\
&T2(2.4)&7 & $5\cdot10^{-6}$ & 0.99999 &0.001579243& [0.001574243, 0.001584243] & 33,772 & 230,258,509,300\\
\hline
\hline
& & & {\bf $\epsilon$} & {\bf $length$} && {\bf Probability interval} & {\bf $CPU_{seq}$} & {\bf $CPU_{par}$}\\
\hline
\multirow{3}{3em}{Prob Reach}& T4(0.6) & 2 & $10^{-9}$ &  8.55e-11 & &[0.0, 8.55e-11] & 52 & 4\\
& T4(1.7) & 6 & $10^{-9}$ & 7.962e-10 & &[9.43986e-08, 9.51948e-08] & 268 & 28\\
& T4(1.8) & 6 & $10^{-9}$ & 9.0e-10 &  &[0.0039559433, 0.0039559442] & 578 & 75\\
\hline
& & & {\bf $\zeta$} & {\bf $c$} & $P$ &{\bf Confidence interval} & {\bf $CPU_{seq}$} & {\bf Sample size}\\
\hline
\multirow{3}{3em}{Monte Carlo}& T4(0.6) & 2 & $5\cdot10^{-6}$ & 0.99999 & 0 & [0, 5e-06] & 32,883 & 230,258,509,300\\
&T4(1.7)&6 & $5\cdot10^{-6}$ & 0.99999 & 9.438088e-08 &[0, 5.094381e-06] & 33,015 & 230,258,509,300\\
&T4(1.8)&6 & $5\cdot10^{-6}$ & 0.99999 & 0.003955074 &[0.003950074, 0.003960074] & 33,354 & 230,258,509,300\\
\hline
\hline
& & & {\bf $\epsilon$} & {\bf $length$} && {\bf Probability interval} & {\bf $CPU_{seq}$} & {\bf $CPU_{par}$}\\
\hline
\multirow{2}{3em}{Prob Reach}& \multirow{2}{2em}{CBB} &2 & $10^{-2}$ & 8.0e-3 & &[0.199, 0.207] & 70 & 15\\
&&2 & $10^{-9}$ & 3.0e-10 & &[0.2049030217, 0.204903022] & 8,332 & 2,581\\
\hline
& & & {\bf $\zeta$} & {\bf $c$} & $P$ &{\bf Confidence interval} & {\bf $CPU_{seq}$} & {\bf Sample size}\\ 
\hline
\noindent{\begin{tabular}{@{}l}
Monte \\ Carlo 
\end{tabular}} & CBB &2 & $5\cdot10^{-3}$ & 0.99 & 0.2045948 &[0.1995948, 0.2095948] & 50,528 & 92,104\\
\hline
\hline
& & & {\bf $\epsilon$} & {\bf $length$} & &{\bf Probability interval} & {\bf $CPU_{seq}$} & {\bf $CPU_{par}$}\\
\hline
\multirow{3}{3em}{Prob Reach}& \multirow{3}{1em}{IG} &1 & $10^{-2}$ & 5.328e-3 & &[0.994589, 0.999917] & 2,805,634 & 165,404\\
&&1 & $10^{-3}$ & 8.1e-4 & &[0.999107, 0.999917] & 3,326,581 & 443,910\\
&&1 & $10^{-4}$ & 5.5e-5 & &[0.999657, 0.999712] & 3,498,765 & 490,257\\ 
\hline
& & & {\bf $\zeta$} & {\bf $c$} & $P$ & {\bf Confidence interval} & {\bf $CPU_{seq}$} & {\bf Sample size}\\
\hline
\multirow{2}{3em}{Monte Carlo}& \multirow{2}{1em}{IG} &1 & $5\cdot10^{-3}$ & 0.99 & 0.997266555 &[0.9945331, 1] & 58,069 & 92,104\\
&&1 & $2.5\cdot10^{-3}$ & 0.99 & 0.99853 &[0.99706, 1] & 219,623 & 368,416\\
\hline

\end{tabular} 
\end{adjustwidth}
\caption{Computing probabilistic reachability with ProbReach and MATLAB.
$k$ = number of discrete transitions; $\epsilon$ = desired size of probability interval;
$length$ = length of probability interval returned by {\em ProbReach}; 
$\zeta,c$ = half-interval width and coverage probability for Chernoff bound;
Sample size = number of simulations (Chernoff bound); $P$ = probability estimate (successes/Sample size);
$CPU_{seq}, CPU_{par}$ = CPU time (sec) of sequential and parallel version; BB = bouncing ball model;
CBB = controlled bouncing ball model; T2(0.6), T2(1.8), T2(2.4) = thermostat model with 2 modes 
at $t = 0.6, 1.8, 2.4$ respectively; T4(0.6), T4(1.7), T4(1.8) = thermostat model with 4 modes 
at $t = 0.6, 1.7, 1.8$ respectively. }
\label{table:results}
\end{table*}

\section{Conclusions and Future Work}
We have presented the {\em ProbReach} tool which computes an arbitrarily small
interval containing the probability that a hybrid system reaches an unsafe region of its
state space. {\em ProbReach} is not limited to a set of predefined random variables, as it works with probability density functions. Thus, it can be extended to support user-defined distributions. 
We have successfully benchmarked {\em ProbReach} and in many cases it demonstrated a better performance in comparison to Monte Carlo simulations while providing stronger guarantees of result correctness. Finally, it was shown that {\em ProbReach} is very efficient for rare event verification.

In the future, we plan to implementing a more efficient parallelisation scheme. This will be performed modifying the partition verification approach. Instead of adding {\em mixed} intervals to a separate queue and verifying them after the main partition, newly partitioned intervals will be pushed to the end of the main queue. Then, a parallelisation manager monitoring the available cores will be dynamically distributing the load equally between the threads, thus reducing CPU idle. According to our estimations, this modification will significantly increase the performance of the tool.

Another extension is to allow probabilistic jumps in the model. We plan to allow jumps whose probabilities
may depend on the (continuous) variables and parameters. 
Finally, we plan to support both nondeterministic and random continuous parameters.
For such systems, probabilistic reachability 
becomes in general an optimisation problem, as the nondeterministic parameters may generate
{\em ranges} of probabilities. These two additions will enlarge very much the class of models analyzable by
{\em ProbReach}.

\section{Acknowledgments}
This work has been supported by award N00014-13-1-0090 of the US Office of Naval Research.

%
\bibliographystyle{abbrv}
\bibliography{/home/paolo/papers/refs}  

\begin{thebibliography}{10}

\bibitem{DBLP:conf/hybrid/AlurCHH92}
R.~Alur, C.~Courcoubetis, T.~A. Henzinger, and P.-H. Ho.
\newblock Hybrid automata: An algorithmic approach to the specification and
  verification of hybrid systems.
\newblock In {\em Hybrid Systems}, volume 736 of {\em LNCS}, pages 209--229,
  1992.

\bibitem{Franzle:STTT14}
C.~Ellen, S.~Gerwinn, and M.~Fr{\"a}nzle.
\newblock Statistical model checking for stochastic hybrid systems involving
  nondeterminism over continuous domains.
\newblock {\em STTT}, 2014.
\newblock To appear.

\bibitem{Sisat}
M.~Fr{\"{a}}nzle, T.~Teige, and A.~Eggers.
\newblock Engineering constraint solvers for automatic analysis of
  probabilistic hybrid automata.
\newblock {\em J. Log. Algebr. Program.}, 79(7):436--466, 2010.

\bibitem{ValidatedIntegration}
S.~Galdino.
\newblock Interval integration revisited.
\newblock {\em Open Journal of Applied Sciences}, 2(4B):108--111, 2012.

\bibitem{DBLP:conf/cade/GaoAC12}
S.~Gao, J.~Avigad, and E.~M. Clarke.
\newblock Delta-complete decision procedures for satisfiability over the reals.
\newblock In {\em IJCAR}, pages 286--300, 2012.

\bibitem{DBLP:journals/corr/GaoKCC14}
S.~Gao, S.~Kong, W.~Chen, and E.~M. Clarke.
\newblock Delta-complete analysis for bounded reachability of hybrid systems.
\newblock {\em CoRR}, arXiv:1404.7171, 2014.
\newblock Available at \url{http://arxiv.org/abs/1404.7171}.

\bibitem{DBLP:conf/cade/GaoKC13}
S.~Gao, S.~Kong, and E.~M. Clarke.
\newblock {dReal}: An {SMT} solver for nonlinear theories over the reals.
\newblock In {\em CADE}, pages 208--214, 2013.

\bibitem{hoeffding}
W.~Hoeffding.
\newblock Probability inequalities for sums of bounded random variables.
\newblock {\em J. Amer. Statist. Assoc.}, 58(301):13--30, 1963.

\bibitem{Hovorka02}
R.~Hovorka et~al.
\newblock Partitioning glucose distribution/transport, disposal, and endogenous
  production during {IVGTT}.
\newblock {\em American Journal of Physiology: Endocrinology and Metabolism},
  282(5):E992 -- E1007, 2002.

\bibitem{filib}
M.~Lerch, G.~Tischler, J.~W.~V. Gudenberg, W.~e. Hofschuster, and
  W.~Kr\"{a}mer.
\newblock {FILIB++}, a fast interval library supporting containment
  computations.
\newblock {\em ACM Trans. Math. Softw.}, 32(2):299--324, 2006.

\bibitem{ADHS09}
P.~J. Mosterman, J.~Zander, G.~Hamon, and B.~Denckla.
\newblock Towards computational hybrid system semantics for time-based block
  diagrams.
\newblock In {\em 3rd IFAC Conference on Analysis and Design of Hybrid Systems
  (ADHS'09)}, pages 376--385, 2009.

\bibitem{Petras:VerifiedNI}
K.~Petras.
\newblock Principles of verified numerical integration.
\newblock {\em Journal of Computational and Applied Mathematics}, 199(2):317 --
  328, 2007.

\bibitem{Sankaranarayanan:2012:SII:2415548.2415569}
S.~Sankaranarayanan and G.~Fainekos.
\newblock Simulating insulin infusion pump risks by in-silico modeling of the
  insulin-glucose regulatory system.
\newblock In {\em CMSB}, volume 7605 of {\em LNCS}, pages 322--341, 2012.

\bibitem{ProbDeltaReach}
F.~Shmarov and P.~Zuliani.
\newblock Verification of probabilistic bounded $\delta$-reachability for
  cyber-physical systems.
\newblock {\em CoRR}, abs/1406.1920, 2014.

\bibitem{SReach}
Q.~Wang, P.~Zuliani, S.~Kong, S.~Gao, and E.~M. Clarke.
\newblock {SReach}: A bounded model checker for stochastic hybrid systems.
\newblock {\em CoRR}, abs/1404.7206, 2014.

\end{thebibliography}
%
%
\appendix
\section{Models}

\subsection{Bouncing ball}
The ball is launched from the initial point $(S_{x} = 0, S_{y} = 0)$ with initial speed 
$\upsilon_0$ which is distributed normally with mean 20 and variance 1, and angle 
to horizon $\alpha = 0.7854$. After each jumps the speed of the ball is reduced by 0.9. The system can be modelled as a hybrid system with one mode with 
dynamics governed by a system of ODEs: 
\begin{equation*}
\begin{split}
S'_{x}(t) &= \upsilon_0 \cos{\alpha} \\
S'_{y}(t) &= \upsilon_0 \sin{\alpha} - gt
\end{split}
\end{equation*}
The goal of the experiment is to calculate the probability of reaching the region 
$S_{x}(t) \ge 100$ within 0, 1, 2 and 3 jumps.

\subsection{Thermostat} 
\begin{figure}[ht!] 
\centering
\includegraphics[width=70mm]{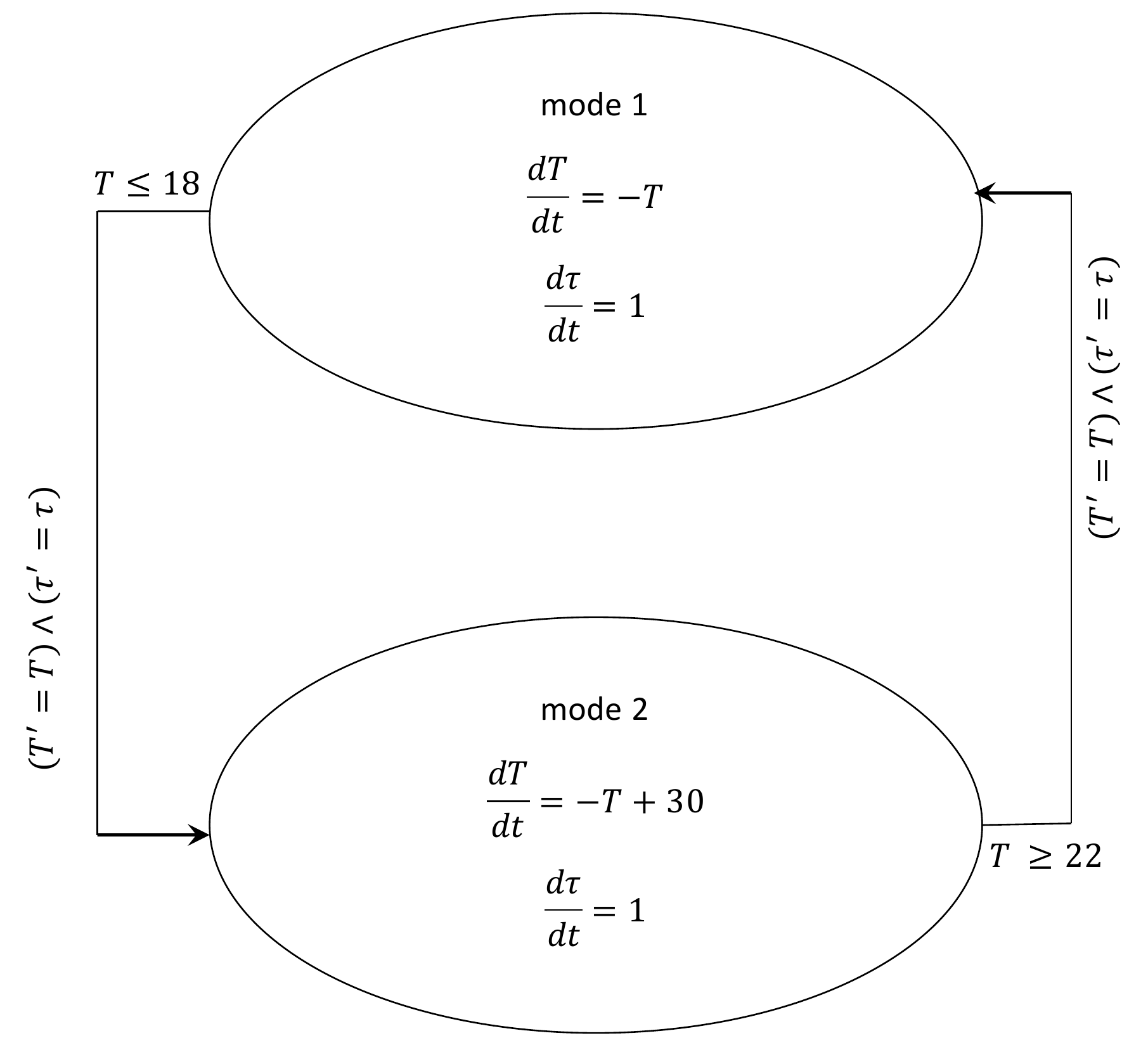}
\caption{A 2-mode thermostat hybrid system}
\label{fig:thermostat-2m}
\end{figure}

The main purpose of the system is to keep the temperature within the desired range. 
The system is modelled by a two mode hybrid system \cite{DBLP:conf/hybrid/AlurCHH92} 
(Fig. \ref{fig:thermostat-2m}). The temperature is changing exponentially and it is 
decreasing in the first mode and increasing in the second mode.
The system starts in mode 1 with the initial temperature $T_{0}$ which is normally 
distributed ($\mu = 30$ and $\sigma = 1$). When the temperature drops to the minimum 
level $T_{min} = 18$, the system makes a transition to mode 2, where the temperature 
increases until it reaches a maximum level $T_{max} = 22$. Then the system makes a 
jump to mode 1 and the loop repeats again. In the model we use function $\tau(t)$ 
to represent the global time, as the current time is reset when the system makes 
a discrete transition.

The goal of the experiment is to calculate the probability of reaching the region 
$T(t) \in [19.9, 20.1]$ in mode 2 at various time points ($\tau = 0.6$, $\tau = 1.8$ and $\tau = 2.4$).

\begin{figure}[ht!] 
\centering
\includegraphics[width=70mm]{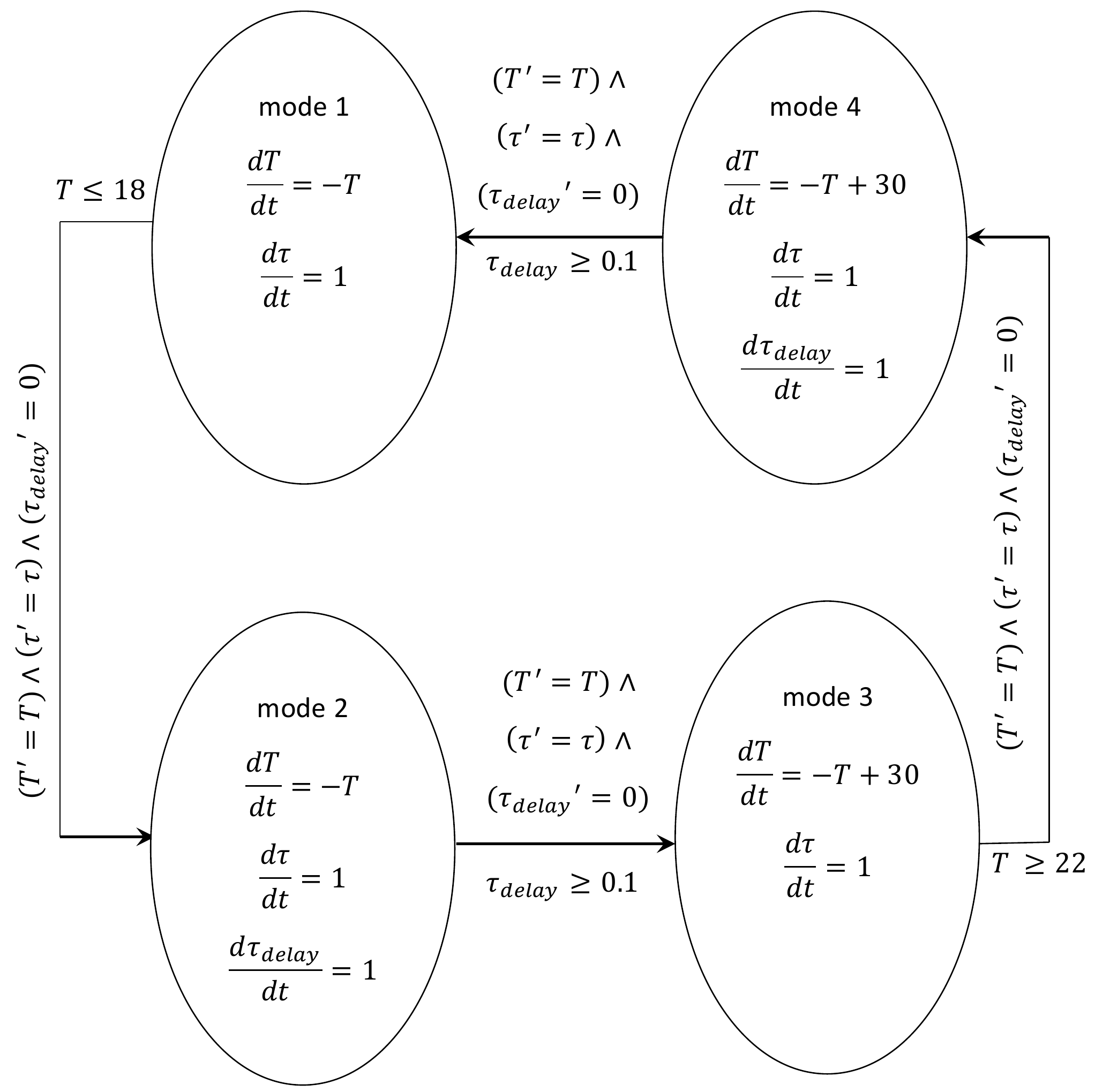}
\caption{A 4-mode thermostat hybrid system}
\label{fig:thermostat-4m}
\end{figure}

We can extend the 2-mode thermostat model to a 4-mode version by adding two 
delay modes (Fig. \ref{fig:thermostat-4m}).
The initial mode of the system is mode 1. Modes 1 and 3 are equivalent to modes 1 and 2 in 
the 2-mode thermostat. Modes 2 and 4 model a delay of 0.1 seconds.
The goal of the experiment is to calculate the probability of reaching the region 
$T(t) \in [19.9, 20.1]$ in mode 3 at various time points ($\tau = 0.6$, $\tau = 1.7$ and $\tau = 1.8$).

\subsection{Controlled bouncing ball}
\begin{figure}[ht!] 
\centering
\raisebox{-0.5\height}{\includegraphics[width=30mm]{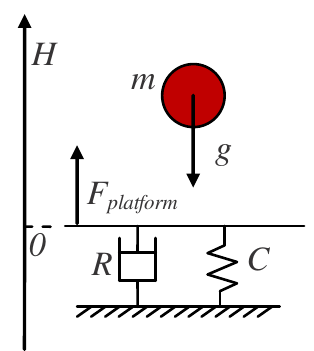}}
\raisebox{-0.5\height}{\includegraphics[width=50mm]{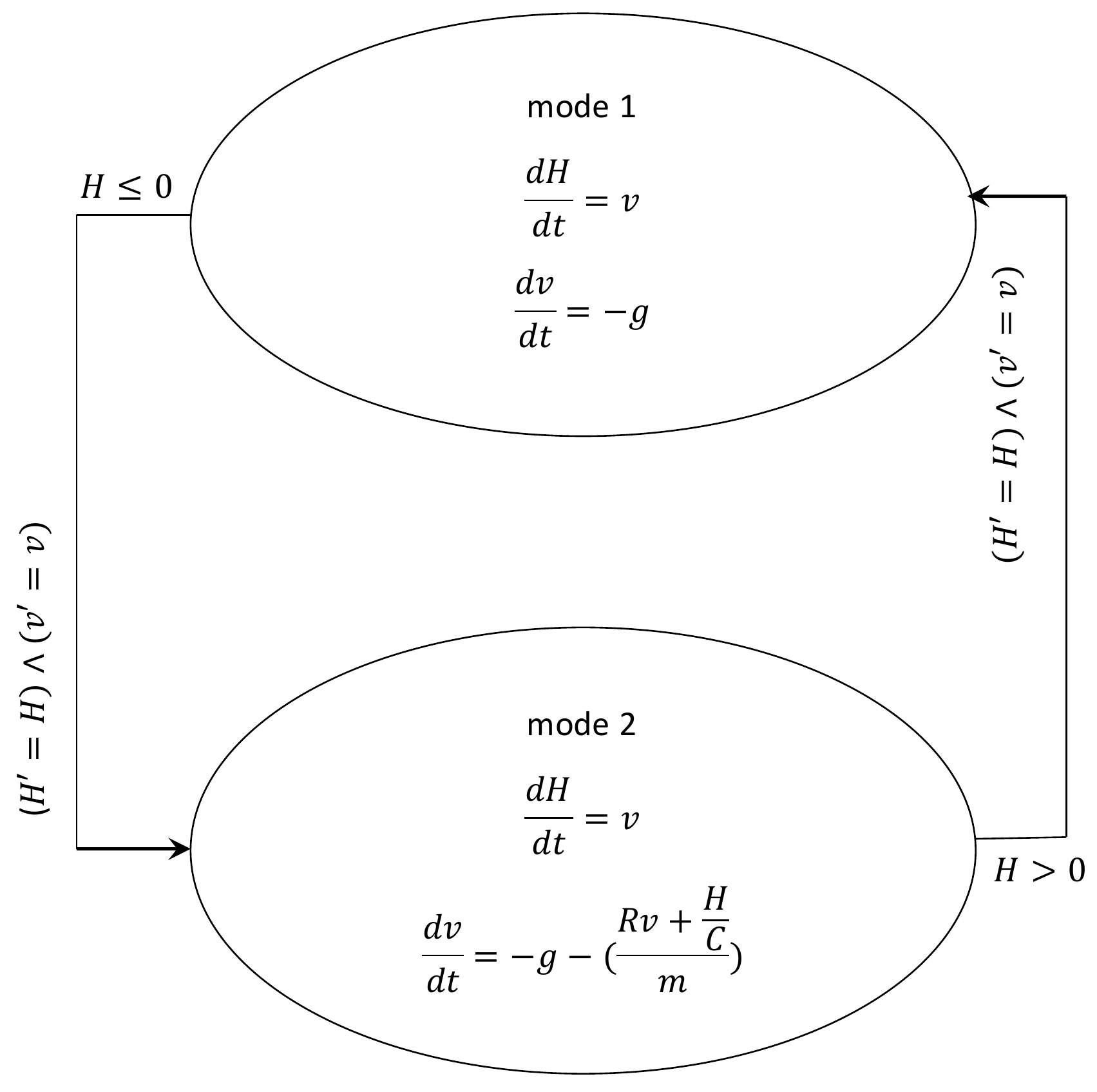}}
\caption{A figure (left hand side) and a model (right hand side) of controlled bouncing ball with $R = 5$, $C = 0.0025$ and $g = 9.8$}
\label{fig:controlled-bouncing-ball}
\end{figure}

Consider a 2-mode hybrid system (Fig. \ref{fig:controlled-bouncing-ball}) modelling a controlled 
bouncing ball \cite{ADHS09}. In mode 1, a ball of mass $m = 7$ is dropped on a platform attached 
to a stiff spring and a damper from a random height $H_{0}$, which is distributed normally 
($\mu = 9$ and $\sigma = 1$). When the ball reaches the platform ($H = 0$) the system makes a 
transition to mode 2, where the ball is reflected from the platform and it jumps back to mode 1 
when the height of the ball is greater than 0.

The goal of the experiment is to calculate the probability that the ball reaches 
the region $H >= 7$ in mode 1 after making one bounce.

\subsection{Insulin-glucose regulatory model}
\begin{figure}[ht!] 
\centering
\includegraphics[width=80mm]{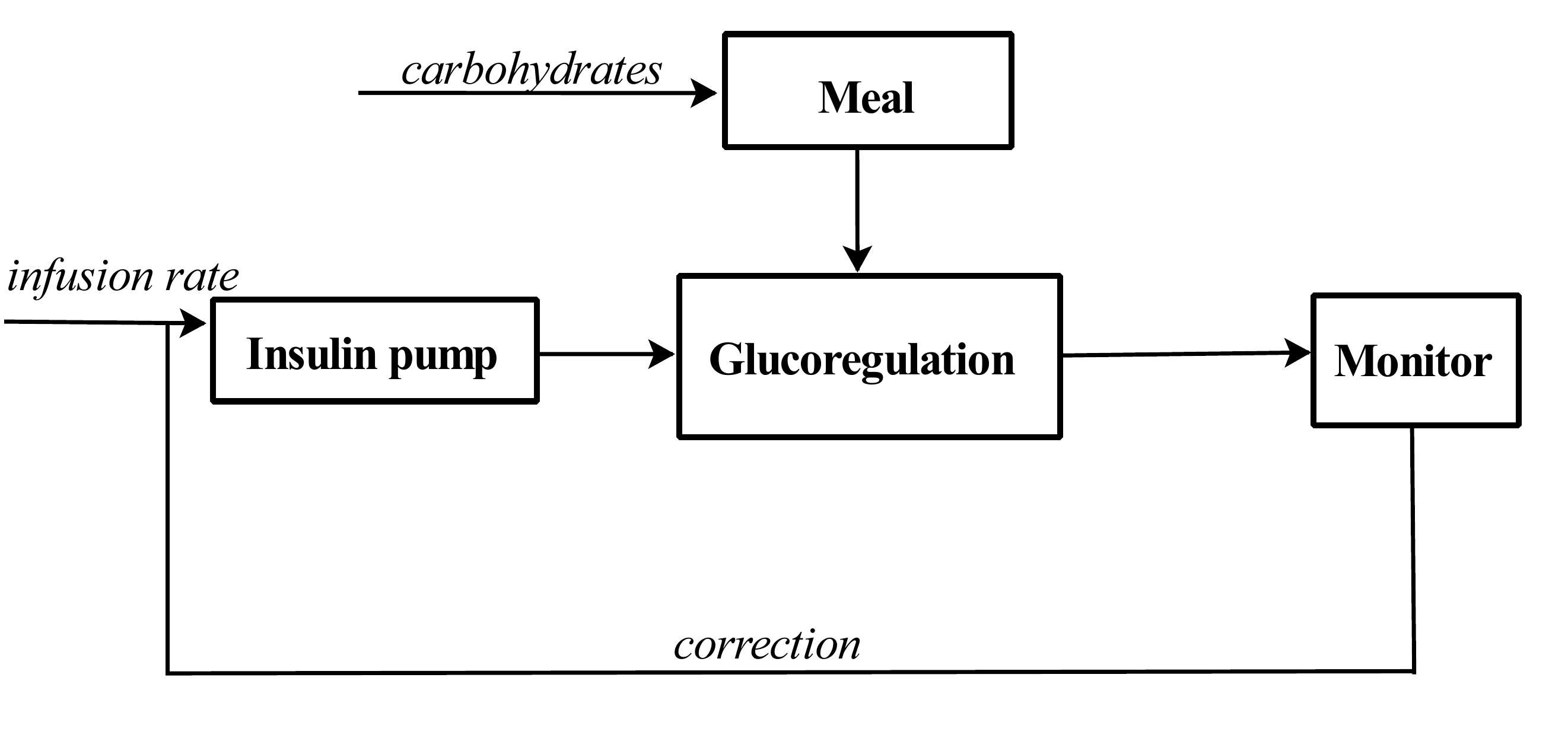}
\caption{Insulin-glucose regulatory high level model}
\label{fig:insulin-infusion-high-level}
\end{figure}

We consider an insulin-glucose regulatory system for the patients with type-1 diabetes. Our tool was applied to the insulin infusion model introduced in \cite{Sankaranarayanan:2012:SII:2415548.2415569} based on Hovorka's glucoregulatory model \cite{Hovorka02}. The system consists of four subsystems: meal, insulin pump, glucoregulatory model and monitor (Figure \ref{fig:insulin-infusion-high-level}). 

The meal parameters (e.g. amount of carbohydrates in the the meal, their glycemic index) are introduced to the meal subsystem by the patient prior to consumption. It models the food absorption by the human guts and outputs a glucose absorption rate. According to the meal characteristics the patient calculates and inputs the initial insulin infusion rate. These two parameters are used by the glucoregulatory subsystem evaluating a glucose level of the patient which is used as an input parameter for the monitor. When the glucose level goes outside the predefined corridor the monitor sends a signal to the pump to increase or decrease the insulin infusion rate. We consider a simplified version the glucoregulatory system assuming that glucose absorption rate ($UG$) and insulin infusion rate ($u_0$) are given explicitly. 
The ODEs of the model are presented in (\ref{eq:insulin-infusion-low-level}), while the model 
parameters and initial conditions are given in Table \ref{table:insulin-glucose-param}.

\begin{equation}\label{eq:insulin-infusion-low-level}
\begin{split} 
\frac{dQ_{1}}{dt} &= -F^{c}_{01} - x_{1}Q_{1} + k_{12}Q_{2}-F_{R}+EGP_{0}(1-x_{3})+ 0.18UG\\
\frac{dQ_{2}}{dt} &=x_{1}Q_{1}-(k_{12} + x_{2})Q_{2}\\
\frac{dS_{1}}{dt} &=u - \frac{S_{1}}{t_{maxI}}\\
\frac{dS_{2}}{dt} &=\frac{S_{1}-S_{2}}{t_{maxI}}\\
\frac{dI}{dt} &=\frac{S_{2}}{t_{maxI}V_{I}}-k_{e}I\\
\frac{dx_{1}}{dt} &=-k_{a1}x_{1}+k_{b1}I\\
\frac{dx_{2}}{dt} &=-k_{a2}x_{2}+k_{b2}I\\
\frac{dx_{3}}{dt} &=-k_{a3}x_{3}+k_{b3}I\\
F_{01}^{c} &=\frac{F_{01}G}{0.85(G+1)}\\
G &=\frac{Q_{1}}{V_{G}}
\end{split}
\end{equation}

\begin{table}[ht!] 
\centering
\begin{tabular}{c c | c c | c c}
\hline\hline 
Param. & Value & Param. & Value & Param. & Value\\ [0.5ex] 
\hline 
$Q_{1}(0)$ & 64.0 & $S_{1}(0)$ & 4.2 & $I(0)$ & 0.03 \\ [0.5ex]
$Q_{2}(0)$ & 40.0 & $S_{2}(0)$ & 4.0 & $x_{1}(0)$ & 0.03 \\ [0.5ex]
$k_{a1}$ & 0.006 & $k_{a2}$ & 0.06 & $k_{a3}$ & 0.03 \\ [0.5ex]
$k_{b3}$ & 0.024 & $k_{e}$ & 0.138 & $k_{12}$ & 0.066 \\ [0.5ex]
$F_{R}$ & 0.0 & $EGP_{0}$ & $0.0161w$ & $u_{0}$ & 0.0 \\ [0.5ex]
$V_{I}$ & $0.12w$ & $V_{G}$ & $0.16w$ & $k_{b2}$ & 0.056 \\ [0.5ex]
$UG$ & 8 & $x_{2}(0)$ & 0.045 & $w$ & 100 \\ [0.5ex]
$t_{max,I}$ & 55 & $u$ & 0.36 & $F_{01}$ & $0.0097w$\\ [0.5ex]
$k_{b1}$ & 0.0034 \\ [0.5ex]
\hline \\ [0.5ex]
\end{tabular} 
\label{table:insulin-glucose-param} 

\caption{Insulin-glucose regulatory model parameters and initial conditions}
\end{table}

The following scenario is considered. Initially a meal is consumed ($UG = 8$) and the insulin pump does not infuse any insulin ($u_{0}=0$). The glucose level starts rising and when it reaches the point of $G = 10$ the monitor sends a signal to the pump to increase the infusion rate ($u = 0.36$). Randomising $x_{3}(0)$ normally ($\mu = 0.05$ and $\sigma = 0.005$) we want to calculate the probability that the glucose level returns back to normal within 60 minutes after the pump started infusion.

\hide{
\section{PDRH files}
This section contains the example of formulas $\phi$ and $\phi^C$ encoding in PDRH format for T2(0.6) model.
\subsection{Formula $\phi$}
\begin{Verbatim}[frame=single, numbers=left, samepage = true]
#define K 1.0
[0, 5] time;
[0, 1000] tau;
//random parameter declaration
N(30, 1) x;
//cooling mode
{ mode 1;
invt:
  (x >= 18);
flow:
  d/dt[x] = - x * K;
  d/dt[tau] = 10.0;
jump:
  (x <= 18) ==> @2 (and (x' = x) (tau' = tau));
}
//heating mode
{ mode 2;
invt:
  (x <= 22);
flow:
  d/dt[x] = - K * (x - 30);
  d/dt[tau] = 10.0;
jump:
  (x >= 22) ==> @1 (and (x' = x) (tau' = tau));
}
//initial state
init:
@1 (and (x >= x_a) (x <= x_b) (tau = 0));
//unsafe region
goal:
@2 (and (x >= 19.9) (x <= 20.1) (tau = 6));
\end{Verbatim}

\subsection{Formula $\phi^{C}$}
\begin{Verbatim}[frame=single, numbers=left, samepage = true]
define K 1.0
[0, 5] time;
[0, 1000] tau;
//cooling mode
{ mode 1;
invt:
  (x >= 18);
flow:
  d/dt[x] = - x * K;
  d/dt[tau] = 10.0;
jump:
  (x <= 18) ==> @2 (and (x' = x) (tau' = tau));
}
//heating mode
{ mode 2;
invt:
  (x <= 22);
flow:
  d/dt[x] = - K * (x - 30);
  d/dt[tau] = 10.0;
jump:
  (x >= 22) ==> @1 (and (x' = x) (tau' = tau));
}
//initial state
init:
@1 (and (x >= x_a) (x <= x_b) (tau = 0));
//unsafe region complement
goal:
@2 (and (or (x < 19.9) (x > 20.1)) (tau = 6));
\end{Verbatim}

}

\end{document}